\documentclass[twocolumn,showpacs,floatfix]{revtex4}
\usepackage{graphicx}

\begin{document}

\title{The spin evolution of spin-3 $^{52}$Cr Bose-Einstein condensate}
\author{Y. Z. He }
\affiliation{State Key Laboratory of Optoelectronic Materials and Technologies, and
Department of Physics, Sun Yat-Sen University, Guangzhou, 510275, P.R. China}

\begin{abstract}
The spin evolution of a Bose-Einstein condensate starting from a
mixture of two or three groups of $^{52}$Cr (spin-$3$) atoms in an
optical trap has been studied theoretically. The initial state is so
chosen that the system does not distinguish up and down. In this
choice, the deviation caused by the single-mode approximation is
reduced. Moreover, since the particle number is given very small
($N=20$), the deviation caused by the neglect of the long-range
dipole force is also reduced.  Making use of these two
simplifications, a theoretical calculation beyond the mean field
theory is performed. The numerical results are help to evaluate the
unknown strength $g_0$.
\end{abstract}


\maketitle

\section{Introduction}

Bose-Einstein condensate (BEC) of atoms with nonzero spin has
greatly attracted the interest of both the experimentalists and
theorists~\cite{HTL1998,OT1998,SDM1998,SJ1998,LCK1998,GA2003,LM2007}
in recent years. Four years ago, the Bose gas of $^{52}$Cr atoms
with electronic spin $J=3$ and nuclear spin $I=0$ was condensed
successfully~\cite{GA2005}. Experimentally, the atomic spins of
$^{52}$Cr were frozen in magnetic trap, but freed in optical
trap~\cite{DBR2006}. For these spin-3 bosons, the interaction
between two atoms is specified by the strengths $g_{S}$, where
$S=0$, $2$, $4$ and $6$ are the total spins of the pair. All $g_{S}$
except the one for $S=0$ have been determined
experimentally~\cite{SJ2005}. To fully understand the characteristic
of $^{52}$Cr and consider the use of this material in application,
it is important to measure $g_{0}$ with the help of theory. Among
all the rich physics of BEC, one attractive phenomenon is the spin
evolution~\cite{LCK1998,SH2004,CMS2004,CMS2005,PH1999,DRB2006}. It
was found that for $^{87}$Rb and $^{23}$Na, the evolution of the
average populations of spin components sensitively depends on the
strengths of interaction~\cite{LMA2007,CZF2008}. Therefore, the
interaction can be determined (or confirmed) by observing the
evolution. However, for $^{52}$Cr, related experimental data and
theoretical analysis are scarce.

In this paper, we consider the spin evolution of a mixture of two or
three groups of $^{52}$Cr atoms in an optical trap. Each atom of a
group has the same $Z$ component of spin, $\mu$. Magnetic fields are
used in the preparation of these groups, but are cancelled during
the whole evolution. What we are interesting in is the effect of the
unknown strength $g_{0}$ on the evolution. When the condensate is
very dilute and the temperature is very low, the single-mode
approximation (SMA, namely, the spatial wave functions of the atoms
distinct in $\mu$ are approximately considered as the same) has been
used by a number of authors to simplify the calculation. This
approximation has also been adopted here. However, even the above
conditions are satisfied, the SMA might not be good as shown
in~\cite{YS2002}. The validity of the SMA depends on the total
magnetization of the system. In the following discussion, a special
initial condition with total magnetization zero is chosen.
Accordingly, the system does not distinguish up and down. Therefore,
the deviation caused by the SMA is expected to be considerably
reduced.

It is well known that the system of $^{52}$Cr contains the
long-range dipole interaction which is considerably stronger than
those of $^{87}$Rb and $^{23}$Na. The dipole interaction is in
general very weak. However, due to being long-range, the combined
effect would be large if numerous atoms are involved. Besides, the
effect would become larger and larger as the evolution goes on.
Accordingly, we consider only the condensate with much fewer
particles and the early stage of evolution. For this very small
condensate, the effect of dipole force is much smaller and therefore
can be neglected. Furthermore, since the particle number $N$ under
consideration is so small, the mean field theory might not work very
well. Thus a method beyond the mean field theory is used in the
follows.

\section{Hamiltonian and the eigenstates}

For a $^{52}$Cr condensate with $N$ atoms, when the dipole force is
neglected, the interaction between a pair of spin-3 bosons $i$ and
$j$ is denoted as $U_{ij}=O_{ij}\delta
(\vec{\mathbf{r}}_{i}-\vec{\mathbf{r}}_{j})$, where
$O_{ij}=g_{0}P_{0}+g_{2}P_{2}+g_{4}P_{4}+g_{6}P_{6}$. And $P_{0}$ to
$P_{6}$ are the projection operators of the $S$-channels. Based on
the SMA, each boson has the same spatial state $\phi
(\vec{\mathbf{r}}_{k})$. When integration over the spatial degrees
of freedom has been performed, the Hamiltonian reads
\begin{equation}
 H
 =\sum_{i<j}U_{ij}
 =f\sum_{i<j}O_{ij},\ \ \
 f=\int |\phi(\vec{\mathbf{r}} )|^{4}d\vec{\mathbf{r}}.
\end{equation}
After the integration, only the spin degrees of freedom are left in
the Hamiltonian. To diagonalize the Hamiltonian, we use the
following Fock-states as the basis functions
\begin{equation}
 |\alpha \rangle
 =|N_{\bar{3}}N_{\bar{2}}N_{\bar{1}}N_{0}N_{1}N_{2}N_{3} \rangle,
 \label{e2_Phi}
\end{equation}
where $\alpha$ indicates the set $\{N_{\bar{3}},\cdots ,N_{3}\}$,
$N_{\mu}$ is the number of bosons with spin component $\mu$, and
$\bar{\mu}$ means $-\mu$. There exist two restrictions on the
$N_{\mu}$ as follows.
\begin{equation}
 \left\{
 \begin{array}{lll}
  N & = & N_{\bar{3}}+N_{\bar{2}}+N_{\bar{1}}+N_{0}+N_{1}+N_{2}+N_{3} \\
  M & = & -3N_{\bar{3}}-2N_{\bar{2}}-N_{\bar{1}}+N_{1}+2N_{2}+3N_{3}
 \end{array}
 \right.,
 \label{e3_NM}
\end{equation}
where $M$ is the total magnetization (a constant). In the Fock
space, by using the factional parentage coefficient~\cite{BCG2004},
the matrix elements of Hamiltonian read
\begin{widetext}
\begin{eqnarray}
 \langle\beta|H|\alpha\rangle
 &=& \langle N'_{\bar{3}} N'_{\bar{2}} N'_{\bar{1}} N'_0 N'_1 N'_2 N'_3
     |f \sum_{i<j} O_{ij}|
     N_{\bar{3}} N_{\bar{2}} N_{\bar{1}} N_0 N_1 N_2 N_3\rangle \nonumber \\
 &=& 2f
  \sum_{\mu\leq\nu;\mu'\leq\nu'} \delta_{\mu'+\nu',\mu+\nu}
  \sum_{S=0,2,4,6} g_S C_{3\mu',3\nu'}^{S,(\mu'+\nu')}
     C_{3\mu,3\nu}^{S,(\mu+\nu)} \nonumber \\
  &&[\ \delta_{\mu'\nu'} \delta_{\mu\nu} \frac{N_{\mu} (N_{\mu}-1) }{4}
   \delta_{\cdots, N_{\mu}-2, \cdots}^{\cdots, N'_{\mu}-2, \cdots} \nonumber \\
  &&+(1-\delta_{\mu'\nu'})\delta_{\mu\nu}\frac{\sqrt{N'_{\mu'} N'_{\nu'} N_{\mu} (N_{\mu}-1) }}{2}
   \delta_{\cdots, N_{\mu}-2, \cdots}^{\cdots, N'_{\mu'}-1, \cdots, N'_{\nu'}-1, \cdots} \nonumber \\
  &&+(1-\delta_{\mu\nu})\delta_{\mu'\nu'}\frac{\sqrt{N_{\mu} N_{\nu} N'_{\mu'} (N'_{\mu'}-1)}}{2}
  \delta_{\cdots, N_{\mu}-1, \cdots, N_{\nu}-1, \cdots}^{\cdots, N'_{\mu'}-2, \cdots} \nonumber \\
  &&+(1-\delta_{\mu\nu})(1-\delta_{\mu'\nu'})\sqrt{N_{\mu} N_{\nu} N'_{\mu'} N'_{\nu'}}
   \delta_{\cdots, N_{\mu}-1, \cdots, N_{\nu}-1, \cdots}^{\cdots, N'_{\mu'}-1, \cdots, N'_{\nu'}-1, \cdots}\ ],
 \label{e4_HMM}
\end{eqnarray}
\end{widetext}
where $C_{3\mu ,3\nu }^{S,(\mu +\nu )}$ is the Clebsch-Gordan
coefficient. In the label $\delta _{\cdots ,N_{\mu}-2,\cdots
}^{\cdots ,N'_{\mu '}-1,\cdots ,N'_{\nu '}-1,\cdots }$, the
superscript denotes a revised set of $\{N'_{\bar{3}},\cdots
,N'_{3}\}$ by reducing both $N'_{\mu '}$ and $N'_{\nu '}$ by 1, and
the subscript denotes a revised set of $\{N_{\bar{3}},\cdots
,N_{3}\}$ by reducing $N_{\mu}$ by $2$. When the two revised sets
are one-to-one identical, the label is $1$, otherwise it is zero.

When both $N$ and $M$ are given, the dimension of the matrix is
finite. After the Hamiltonian is diagonalized, the $j$-th
eigenenergy $E_{j}$ and the corresponding eigenstate $\psi _{j}$ are
obtained. $\psi _{j}$ can be expanded by the basis functions (or
vice versa) as follows.
\begin{eqnarray}
 \psi _{j}=\sum_{\alpha}c_{\alpha}^{j}|\alpha \rangle
 \ \ \ \mbox{or}\ \ \
 |\alpha \rangle=\sum_{j}c_{\alpha}^{j}\psi _{j},
\end{eqnarray}
where the coefficient $c_{\alpha}^{j}=\langle \alpha |\psi
_{j}\rangle$ is real. These eigenstates are used in the following
description of evolution.

\section{Evolution of population of spin components}

The spin evolution begins when the three groups of atoms are mixed
together. All the atoms of the first group have $\mu =3$, those of
the second have $\mu =-3$ and those of the third have $\mu =0$.
Therefore, the initial state is just a Fock-state $|I\rangle
=|N_{\bar{3}},0,0,N_0,0,0,N_{3}\rangle$, where $N_{3}$,
$N_{\bar{3}}$, and $N_{0}$ are the number of atoms in the first to
third groups, respectively. $|I\rangle$ can be expanded by the
series of $\psi_{j}$ with the coefficients $c_{I}^{j}$. Thus the
time-dependent solution of the Schr\"{o}dinger equation $\Psi (t)$
describing the evolution reads
\begin{eqnarray}
 \Psi(t)
 =e^{-iHt/\hbar }|I\rangle
 =\sum_{j}c_{I}^{j}e^{-iE_{j}t/\hbar}\psi _{j}.
\end{eqnarray}
Since $c_{I}^{j}$ are known constants determined by $|I\rangle$,
$E_{j}$ and $\psi _{j}$ are also known. So the evolution can be
fully understood.

From $\Psi (t)$, we define the time-dependent population $P_{\mu
}^{I}(t)$ which is the probability of an atom in spin component
$\mu$ at $t$. It reads,
\begin{equation}
 P_{\mu}^{I}(t)
 =\frac{1}{N}\langle \Psi (t)|a_{\mu}^{+}a_{\mu}|\Psi(t)\rangle
 =B_{\mu}^{I}+O_{\mu}^{I}(t)
 \label{e8_Pu}
\end{equation}
where $a_{\mu}^{+}$ and $a_{\mu}$ are the creation operator and
annihilation operator of an atom in $\mu$, respectively.
\begin{eqnarray}
 \label{e10_Ou}
 B_{\mu}^{I}
 &=&\sum_{j}
    (c_{I}^{j})^{2}
    \sum_{\alpha}
    (c_{\alpha}^{j})^{2}
    \frac{N_{\mu}^{\alpha}}{N} \\
 O_{\mu}^{I}(t)
 &=&2\sum_{j<j'}
    \cos [\frac{(E_{j'}-E_{j})t}{\hbar }]
    c_{I}^{j}c_{I}^{j'}
    \sum_{\alpha}
    c_{\alpha}^{j}c_{\alpha}^{j'}
    \frac{N_{\mu}^{\alpha}}{N}
\end{eqnarray}
$N_{\mu}^{\alpha}$ is the number of atoms in $\mu$ within $|\alpha
\rangle$. Equation~(\ref{e8_Pu}) contains two terms. The first one
$B_{\mu}^{I}$ is only determined by $|I\rangle$ and is
time-independent and therefore, appears as a background of
oscillation. The second one $O_{\mu}^{I}(t)$ contains the
time-dependent factor $\cos [(E_{j^{\prime }}-E_{j})t/\hbar ]$ which
implies an oscillation upon the above background. At the beginning
of evolution (\textit{i.e.}, $t=0$), $\Psi (0)=|I\rangle$ and
$P_{\mu}^{I}(0)=N_{\mu}^{I}/N$. Moreover, if the initial state has
the symmetry $N_{\bar{\mu}}^{I}=N_{\mu}^{I}$, the population would
also be symmetric where $P_{\bar{\mu}}^{I}(t)=P_{\mu}^{I}(t)$. This
is because in this case, the $Z$ axis can be reversed.

In the following calculation, $N=20$ is given. The initial states
are given as $N_{\pm 3}^{I}=(N-N_{0}^{I})/2$ and $N_{\pm
2}^{I}=N_{\pm 1}^{I}=0$, where $N_{0}^{I}$ is even and is ranged
from $0$ to $N$. In this choice, $|I\rangle$ is uniquely determined
by $N_{0}^{I}$. Obviously, the system has the up-down symmetry. And
the total magnetization $M$ is zero, which is a condition in favor
of the SMA. The $10^{-8}$~meV, ${\mathring{A}}$ and $\sec$ are used
as units for energy, length and time, respectively. The strengths
$g_{2}$, $g_{4}$ and $g_{6}$ are taken from~\cite{WJ2005}, namely,
they are $-3.88$, $30.80$ and $59.64,$ respectively, whereas $g_{0}$
will be given at a number of testing values. The average density is
given as $f=2\times 10^{-11}$. This value is simply evaluated under
a model that the density is uniform inside a sphere. A slight
deviation of $f$ does not affect the following qualitative results.

\subsection{The background $B_{\mu}^{I}$}

\begin{figure}[htbp]
 \centering
 \resizebox{0.95\columnwidth}{!}{\includegraphics{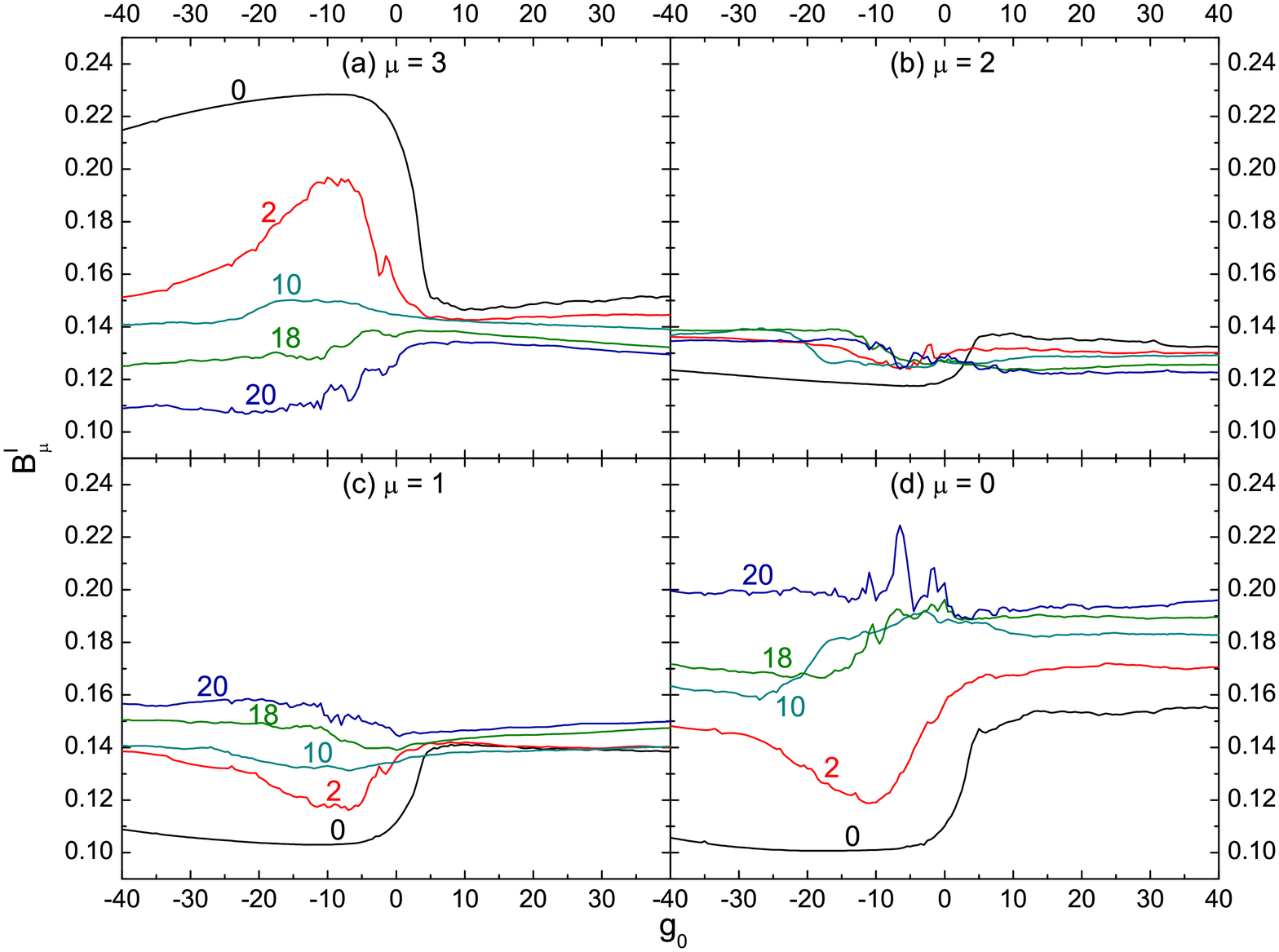}}
 \caption{(Color online.) The background of spin evolution,
$B^{I}_\mu$, against $g_0$ where (a)~$\mu=3$, (b)~$\mu=2$,
(c)~$\mu=1$ and (d)~$\mu=0$, respectively. Five cases of initial
states where $N_0^I=0$, $2$, $10$, $18$ and $20$, respectively, are
given and marked by the curves.}
 \label{crfig1}
\end{figure}

It is proved that the background of oscillation in spin-evolutions
of spin-1 condensates does not depend on the interaction
\cite{LB2008}. However, the argument of that paper is based on the
uniqueness of the Fock-state with a given $N$, $M$, and $N_{0}$.
Obviously, the uniqueness holds no more for spin-3 systems.
Therefore, the knowledge of interaction might generally be obtained
by observing $B_{\mu}^{I}$. This is shown in Fig.~\ref{crfig1} where
the dependency on the initial state and on $g_{0}$ is revealed. It
is also shown that $B_{\mu}^{I}$ would depend on $g_{0}$ rather
weakly if $g_{0}$ is positive. In this case, the structures of
low-lying eigenstates would be dominated by $S=2$ pairs (because
only in this kind of pairs, the two atoms are mutually attracted).
However, $B_{\mu}^{I}$ would depend on $g_{0}$ rather strongly if
$g_{0}$ is negative and close to $g_{2}$. In this case, the
structures of the eigenstates would vary sensitively with $g_{0}$
due to the competition of the $S=2$ and $S=0$ pairs~\cite{BCG2009}.
As a result, there is a domain of sensitivity. If the realistic
$g_{0}$ turns out to fall in this domain, it could be determined by
observing $B_{\mu}^{I}$.

The background can be rewritten as $B_{\mu}^{I}=\sum_{j}W_{j}^{I}
Q_{\mu}^{j}$, where $W_{j}^{I}=(c_{I}^{j})^{2}$ is the weight of
$\psi _{j}$ in $|I\rangle $, $Q_{\mu}^{j}=\sum_{\alpha}(c_{\alpha
}^{j})^{2}N_{\mu}^{\alpha}/N \equiv \langle \psi
_{j}|a_{\mu}^{+}a_{\mu}|\psi_{j}\rangle /N$ is the probability of an
atom in $\mu$ within $\psi _{j}$. Note that the curve with
$N_{0}^{I}=0$ is much higher in Fig.~\ref{crfig1}(a), but much lower
in Fig.~\ref{crfig1}(d). In this state, all the spins are either up
($\mu =3$)\ or down ($\mu =-3$) initially. Therefore, those $\psi
_{j}$ with a larger $Q_{\pm 3}^{j}$ (\textit{i.e.}, having averagely
more atoms for $\mu =\pm 3$) would have a larger weight $W_{j}^{I}$.
This fact explains why the curve with $N_{0}^{I}=0$ is the highest
in Fig.~\ref{crfig1}(a), where the $\mu =3$ atoms are observed.
Meanwhile, those $\psi _{j}$ with a larger $Q_{0}^{j}$ would have a
smaller weight\ $W_{j}^{I}$, which explains why the curve with
$N_{0}^{I}=0$ is the lowest in Fig.~\ref{crfig1}(d).

\subsection{The oscillation $O_{\mu}^{I}(t)$}

\begin{figure}[htbp]
 \centering
 \resizebox{0.95\columnwidth}{!}{\includegraphics{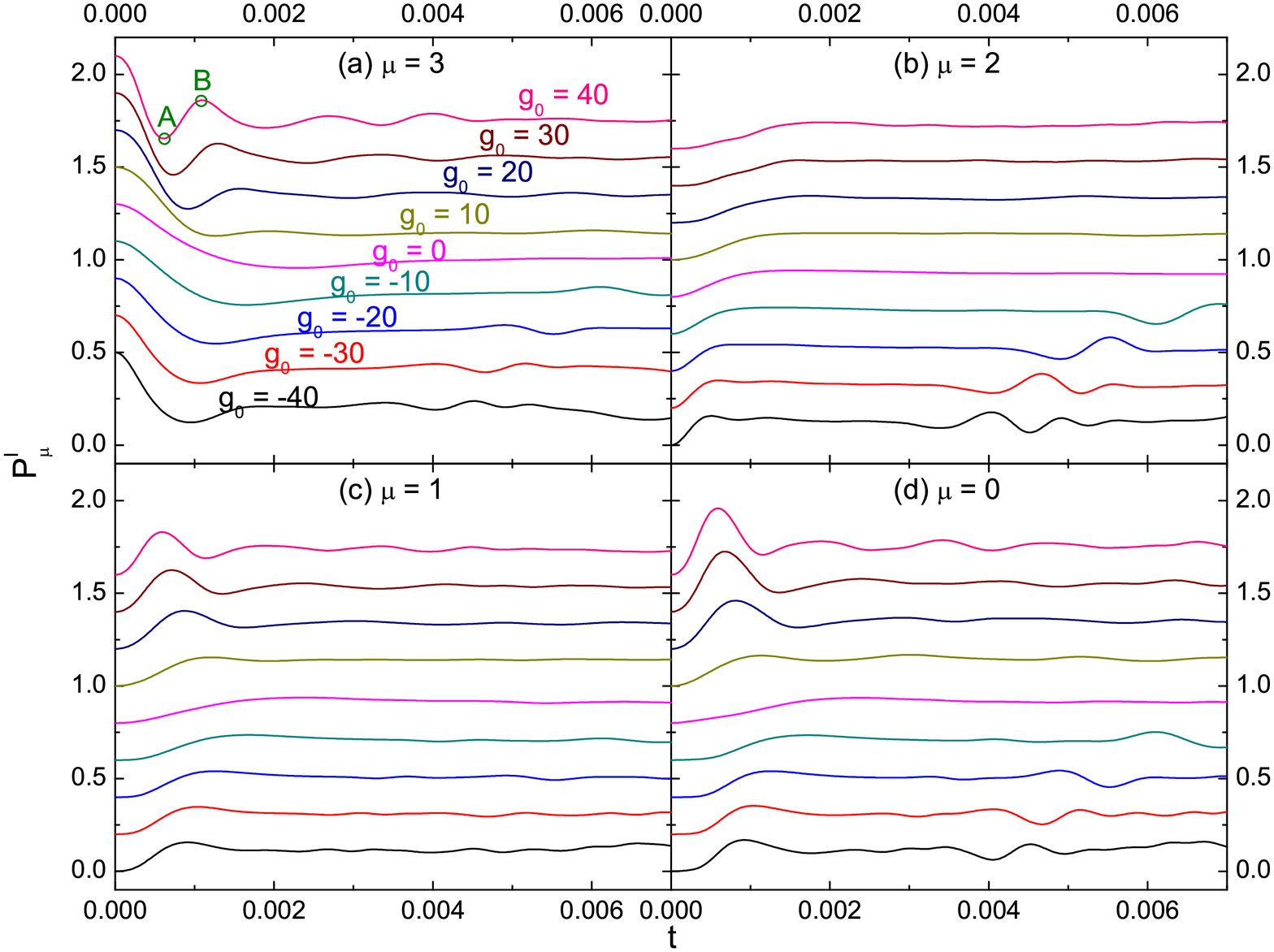}}
 \caption{(Color online.) The populations of spin evolution,
$P^{I}_\mu(t)$, against $t$ (in second) where (a)~$\mu=3$,
(b)~$\mu=2$, (c)~$\mu=1$ and (d)~$\mu=0$, respectively. The initial
state has $N_0^I=0$. In every subfigure, the curves from the lowest
to the highest have $g_0$ from $-40$ to $40$ with a step $10$ as
marked in (a). Each curve has been shifted upward by $0.2$ more than
its lower adjacent to guide the eyes.}
 \label{crfig2}
\end{figure}

\begin{table}[htbp]
\caption{The first minimum and the second maximum of $P^{I}_{3} (t)$
denoted as $P_{3,A}^I$ and $P_{3,B}^I$, respectively, for $N_0^I=0$.
$T_{3,A}^I$ and $T_{3,B}^I$ are their locations (in ms).}
\begin{ruledtabular}
 \label{crtab1}
 \begin{tabular}{lrrrrrrrrrrrr}
 $g_0$        & $-40$  & $-30$  & $-20$  & $-10$  & $0$    & $10$   & $20$   & $30$   & $40$   \\
 \hline
 $P_{3,A}^I$  & $0.12$ & $0.13$ & $0.15$ & $0.16$ & $0.16$ & $0.13$ & $0.07$ & $0.06$ & $0.05$ \\
 $T_{3,A}^I$  & $0.94$ & $1.08$ & $1.26$ & $1.66$ & $2.30$ & $1.26$ & $0.92$ & $0.74$ & $0.60$ \\
 $P_{3,B}^I$  & $0.21$ & $0.21$ & $0.25$ & $0.25$ & $0.21$ & $0.15$ & $0.18$ & $0.23$ & $0.26$ \\
 $T_{3,B}^I$  & $1.84$ & $2.62$ & $4.88$ & $6.10$ & $5.60$ & $1.94$ & $1.58$ & $1.30$ & $1.10$
 \end{tabular}
 \end{ruledtabular}
\end{table}

In Eq.~(\ref{e10_Ou}), the time factor $(E_{j^{\prime
}}-E_{j})/\hbar$ is in general not a multiple of integer among all
pairs of $j$ and $j'$. Therefore $P_{\mu}^{I}(t)$ is non-periodic
which is illustrated in Fig.~\ref{crfig2}.

In the following discussion, we firstly focus on the case where
$N_{0}^{I}=0$. This case also implies that $P_{\pm 3}^{I}(0)=0.5$ at
the beginning. Afterward, due to the appearance of other components,
$P_{\pm 3}^{I}(t)$ go down from the maximum $0.5$ as shown in
Fig.~\ref{crfig2}a, while others $P_{\mu}^{I}(t)$ go up from 0. And
they fluctuate around the backgrounds.  Let the first minimum of
$P_{\mu}^I (t)$ in Fig.~\ref{crfig2}(a) be denoted as $A$ located at
$t= T_{\mu,A}^I$, and the second maximum in the figure be denoted as
B located at $t=T_{\mu,B}^I$. Related data are given in
Tab.~\ref{crtab1}. It is clearly shown that from the locations of
the maximum and minimum obtained via the theoretical calculation,
the strength $g_0$ can be determined once the realistic locations
are experimentally measured. Additional information can also be
extracted from Figs.~\ref{crfig2}(b) and \ref{crfig2}(d). For
example, the first peak of $P_{\mu}^{I}(t)$ with $\mu = 1$ or $0$
appearing in the earliest stage of evolution can help to
discriminate $g_0$.

\begin{figure}[htbp]
 \centering
 \resizebox{0.95\columnwidth}{!}{\includegraphics{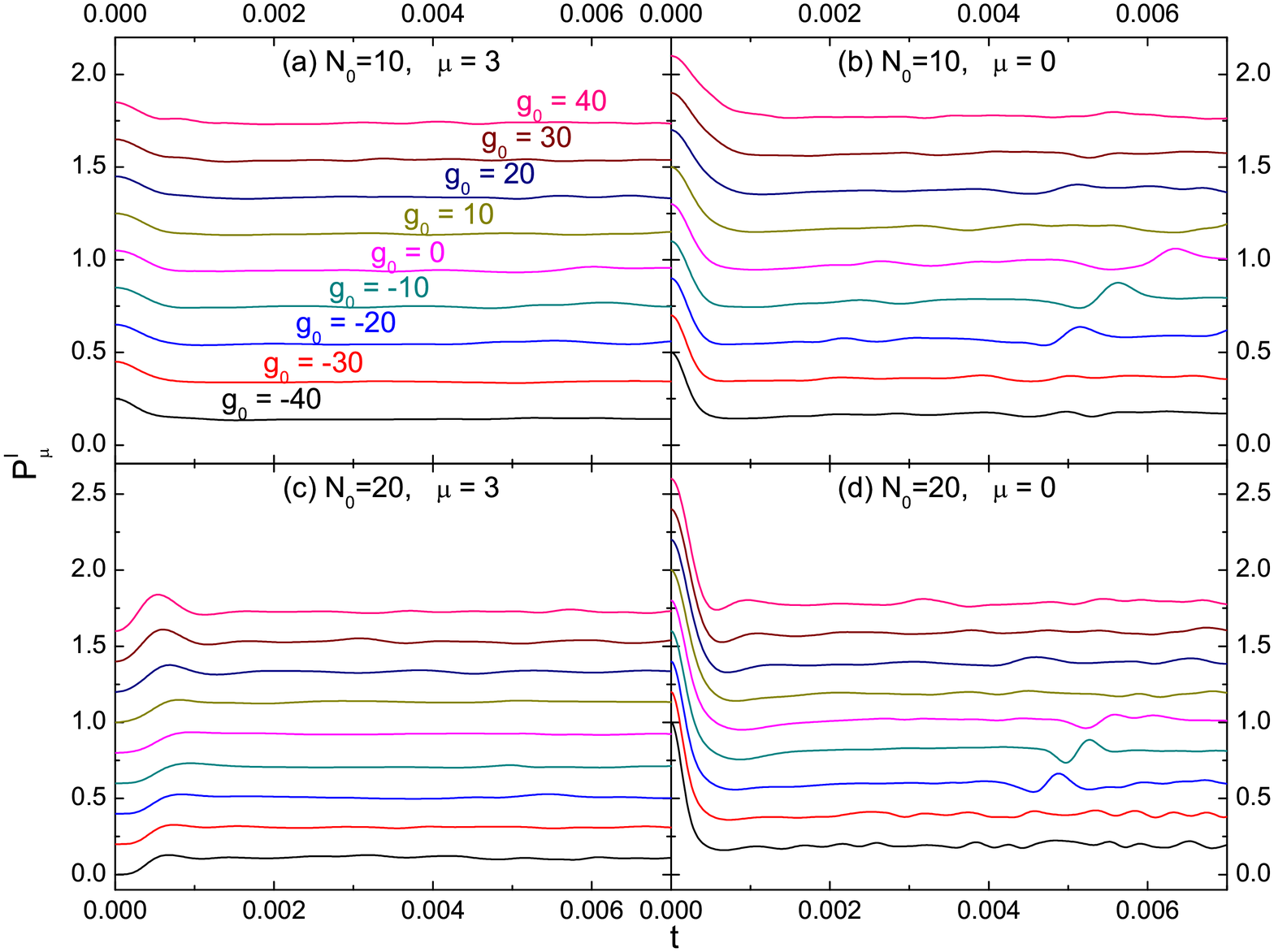}}
 \caption{(Color online.) The same as Fig.~\ref{crfig2} expect
that $N_0^I=10$ for (a)~$\mu=3$ and (b)~$\mu=0$, and $%
N_0^I=20$ for (c)~$\mu=3$ and (d)~$\mu=0$, respectively.}
 \label{crfig3}
\end{figure}

For comparing, the evolutions for both $N_0^I=10$ and $20$ are
illustrated in Fig.~\ref{crfig3}. It is shown that the evolutions
are no more sensitive to $g_0$.  Thus we conclude that $N_0^I=0$ as
shown in Fig.~\ref{crfig2} is a much better choice.

\section{Conclusion}

We study the spin evolution starting from a mixture of two groups of
$^{52}$Cr atoms, which are fully polarized but in reverse directions
and contains only a few particles. And we find an effective way for
determining the strength $g_0$. In this way, the deviations caused
by the SMA and by the neglect of the dipole force are reduced.
Accordingly, the theoretical approach becomes much simpler and a
calculation beyond the mean field theory is performed. The numerical
results show that the knowledge on $g_0$ can be thereby extracted.
Nonetheless, the above theoretical calculation can only provide a
rough evaluation of $g_0$. For an accurate determination, more
precise theory beyond the SMA and with the dipole force taking into
account is necessary. This will lead to a great complexity, and
hopefully can be realized in the near future.

\begin{acknowledgments}
This work is supported by the NSFC under the Grant No. 10874249.
\end{acknowledgments}

\end{document}